# Givers & Receivers perceive handover tasks differently:

## Implications for Human-Robot collaborative system design

Roy Someshwar • Yael Edan

**Abstract:** Human-human joint-action in short-cycle repetitive handover tasks was investigated for a bottle handover task using a three-fold approach: work-methods field studies in multiple supermarkets, simulation analysis using an ergonomics software package and by conducting an in-house lab experiment on human-human collaboration by re-creating the environment and conditions of a supermarket. Evaluation included both objective and subjective measures. Subjective evaluation was done taking a psychological perspective and showcases among other things, the differences in the way a common joint-action is being perceived by individual team partners depending upon their role (giver or receiver). The proposed approach can provide a systematic method to analyze similar tasks. Combining the results of all the three analyses, this research gives insight into the science of joint-action for short-cycle repetitive tasks and its implications for human-robot collaborative system design.

## 1. INTRODUCTION

What is one thing common among – drill sessions, military parade and professional rowing? They are all examples of short-cycle and repetitive joint-actions among humans. According to Sebanz et. al. [1] "joint action can be regarded as any form of social interaction whereby two or more individuals coordinate their actions in space and time to bring about a change in the environment". Synchronized joint-action is commonly observed among professionals and experts, however, knowingly or unknowingly [2] every human coordinates his/her actions with the other all the time, be it during aerobic classes or while doing the dishes with a partner.

---

Roy Someshwar (✉) • Yael Edan
Dept. of Industrial Engineering & Management,
Ben-Gurion University of the Negev, Beer Sheva 84105, Israel
E-mail: royso@bgu.ac.il

The science of joint-action is primarily an area of cognitive psychology [3], [4], philosophy [5] and musicology [6][7]. Recently, this field has received much attention in the Human-Robot Interaction (HRI) community [8]–[10]. This is because, the results of human-human joint action is of profound importance for designing friendlier and ergonomic Human-Robot (H-R) collaborative systems and such design methodology and principles have been successfully implemented in [11]–[15]. The worldwide popularity of the toy robot Keepon [16] is a simple example of designing human-friendly robots by taking cues from human-human rhythmic interaction.

A joint-action may involve direct or indirect handover between the collaborating partners. Studies in human-human handover, e.g., [12], [17]–[20] have been instrumental in designing robots with human like handovers, e.g., [17], [21], [22]. These human-human and human-robot handovers generally deal with face-to-face handover scenarios where communication between partners is possible through eye-gaze and other non-verbal communication cues [21]. In such cases, the point-of-handover (p-o-h) in a joint-action is generally determined by the giver [19]. In short-cycle repetitive handovers, however, there may or may not exist an eye-gaze in every handover cycle. In the absence of an eye-gaze by the receiver, a-priori expectation of the receiver about the probable p-o-h plays a more significant role in the success of the handover [18].

Investigation in a goal-directed joint-task [11] (like e.g., moving a table together) shows that human tend to synchronize their arm movements, which in turn, calls for precise movements. In general, movement synchronization is a guiding dynamical process which leads to stable coordination patterns in natural human-human joint action [9]. Hence, synchronization among the team partner must exist to be able to work together. This helps in building team coordination which plays an important role in joint-action in repetitive tasks.

The fluency [23] of team coordination depends on many implicit and explicit factors, including team communication [24], [25], agreeableness among team members [26], [27], habit persistence for their preferences [28] and their ability to adapt, attend and anticipate [29]. Adding to that, a joint-action in short-cycle repetitive task involves closer and more frequent interaction. An ergonomic and friendlier H-R system is thus critical to its success if a robot is expected to collaborate in such a joint-action.

The contribution of this article is a systematic approach to investigate human–human joint action in short-cycle repetitive tasks in a quantitative and qualitative way. It helps in developing an understanding of the influencing parameters of such joint-actions and hence its possible implications for designing future human-robot collaborative systems. Such collaborative systems can be very useful in executing efficiently the pick-and-place and packing related tasks together with human in future supermarkets and warehouses equipped with mobile robot systems for logistics [30].

The study was performed by taking a three-fold approach – (a) work-methods analyses of field studies on supermarket workers were carried out by visiting three supermarket chains multiple times; (b) work-methods and ergonomic analysis of the same task in a virtual simulation environment; and, (c) an in-house lab experiment on human-human joint-action performed by re-creating the environment and conditions of a supermarket. The focus was on the last step since validation in real-world conditions is critical.

In our research on joint-action, we chose a typical collaborative task which is short-cycle and repetitive in nature – replenishing soft-drink bottles of 1.5 liters in the store shelves. Replenishing products in shelves is, in fact, a commonly observed task in supermarkets. During our visits to the supermarket, we observed that this activity picks up pace during rush-hours and/or prior to holiday seasons. From the field study, it can be said that one individual cycle of this task generally varies between 1 to 4 seconds depending upon the complexity of the exact given task, shelf height and given demand (pace) and hence it can be classified as a short-cycle repetitive task corresponding to the literature [31]–[34]. A short-cycle repetitive task is defined as a physical task done by a human with an individual cycle length of the task / sub-task varying approximately between 2 sec (or less) to a maximum of 20 sec [31], [32].

Combining the results of all the three analyses, this research gives insights into the science of joint-action for short-cycle repetitive tasks. This work is part of a larger project focusing on developing a framework for H-R collaborative system for industrial settings. A set of H-R system design implications is an outcome of the current work which were used in developing three human-robot collaboration models, namely, Timing based, Sensor based and Adaptive model. These models were tested, validated and compared through analytical and simulation analyses and experiments reported in [35]–[38].

## 2. METHODS

The human-human joint-action in short-cycle repetitive handover tasks was analyzed through a real-world case-study. The scenario investigated represents a typical job of supermarket workers – the task of stacking bottles in store shelves from the cartons. The joint-action investigated was the hand-over of bottles in a supermarket. The selection of a real-world task is important for providing reliable results and even more, its analyses in real-world conditions is essential for valid results.

From the field studies, we observed that this job is generally done by two people together as a team, each with a specific role. One is a giver whose job is to pick up a bottle from the carton and hand it over. The other is a receiver whose job is to collect this bottle from the hands of a giver and place (and align) it at the right location in the shelf. Below, each of the three methods used to study this joint-action has been explained.

Analyses were conducted using three methods – (a) Field studies of work-methods in supermarkets; (b) Ergonomic simulation analysis using Jack software; (c) In-house lab experiments by re-creating the conditions of a supermarket. The measures employed for each of the above methods are discussed in detail in the sections below.

The motivation behind the work-methods field studies was to get a first-hand understanding of the scenario, the problem, the needs, nature of the task and at the same time, to study the work-methods of humans working in teams in supermarkets. It was carried out by visiting three supermarkets of different chains at different locations in Israel and recording data in real world conditions followed by data analyses.

The software simulation study was conducted to analyze the involved bio-mechanics and ergonomic aspects of short-cycle repetitive handover tasks. Although this is not the main focus of this paper, it provided an understanding which of the sub-tasks is physically more strenuous and hence could be delegated to a robot if a human-robot collaborative system is commissioned for the given job.

Finally, based on the observations and the data collected from the field-study and simulation analyses, the set-up of the in-house lab experiment was designed. It also helped in defining the experimental variables and conditions which resulted in re-creating similar conditions as that of a supermarket. The lab experiment was the main focus of the research. It was aimed at investigating the psychological aspect of human-human joint-action from the perspective of a giver and a receiver. All experiments were formally approved by the university's Human Subject Research Committee.

## 3. WORK-METHODS FIELD STUDIES IN SUPERMARKETS

Work-methods analyses were conducted at three different supermarket chains in Israel.

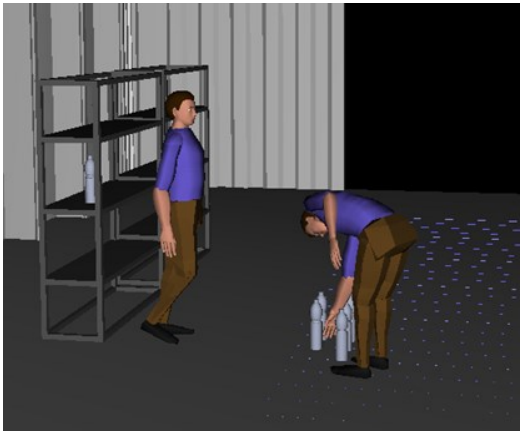 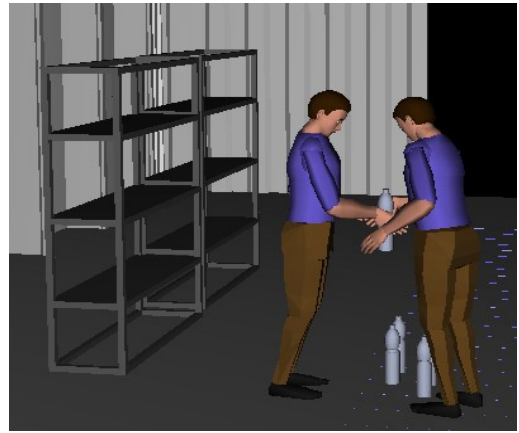

**Fig. 1a & 1b** An example of a short-cycle repetitive handover task in the software simulation environment

A sample video of the field study is presented in Online Resource 1. In most of the locations however, video recording was not allowed by the store manager and in these cases, the following objective data was manually recorded by two observers following worker's consent and knowledge – the cycle time of each handover, number of missed or unsynchronized handovers, the difference in height between the given carton of bottles (upper carton / lower carton) and the shelf, the rest time and the total ON time. The two observers were assigned with different responsibilities. One was responsible for noting the cycle times of each handover alone; the other was responsible for noting the number of missed or unsynchronized handovers and other objective measures as mentioned above. According to [39], ON time in repetitive tasks is defined as the mean time spent working together for one continuous period. Data was recorded separately for the higher, medium and lower shelf.

Table 1 gives a comparative view of the four extreme effort tasks (the first four tasks) with the minimum effort task (the fifth one). Medium shelf - Higher carton [MS-HC] represents the minimum effort task because it required the least amount of lower back bending and other physical movement.

The average cycle time varied between 1.7 sec *(SD = 0.5)* and 3.3 sec *(SD = 1.4)*. The workers performed the fifth task [MS-HC] with the lowest average cycle time. This time is 48% lower than the avg. cycle time for the first task involving Lower shelf - Lower carton (LS-LC).

| Task | Shelf height (cm) | Bottle carton height (cm) | Average Cycle Time (sec) | Std. Deviation (sec) | %age of Synchronized Handovers |
|---|---|---|---|---|---|
| LS-LC | 10 | 40 | 3.3 | 1.4 | 87.51% |
| HS-HC | 165 | 110 | 2.3 | 0.9 | 82.56% |
| LS-HC | 10 | 110 | 3.7 | 1 | 88.73% |
| HS-LC | 165 | 40 | 3.4 | 0.9 | 88.05% |
| MS-HC | 124 | 110 | 1.7 | 0.5 | 76.80% |

**TABLE 1** Results of work-methods field studies in supermarkets. Task: LS / MS / HS = Lower / Mid / Higher Shelf; LC / UC = Lower / Upper Carton

The number of missed or unsynchronized handovers over a single ON time varied between 7 and 19%. Overall, the speed and efficiency of the team was found to be dependent upon the relative difference of height between the given carton of bottles and the shelf which, in turn, defines the amount of relative bending required for the task. The shelves in the supermarket were of the dimensions: 165 cm – the upper shelf height, 124 cm – the medium shelf height and 10 cm – the lower shelf height. The lower/upper carton height (LC/UC) were 40 cm and 110 cm respectively (floor is the reference plane of these measurements).

## 4. SOFTWARE SIMULATION

Jack 8.0.1 [40] is a simulation software package from Siemens used for modeling humans in workplace environments aimed to address the ergonomic aspects of manual operations. It is commonly used for testing and validating designs and operations for a wide variety of human factors, including injury risk, fatigue limits, time of operation, user-comfort, line-of-sight, energy expenditure and other important parameters [40].

The tool *Task Simulation Builder* in the simulation software accepts high-level commands to instruct the human model in a virtual work environment. This capability facilitates quick animation and scenario development. Once a given task sequence is explicitly defined for the human model, different what-if scenarios can be tested and evaluated. This is done by swapping in human figures of different physical attributes, by moving objects in the environment or by changing the heights and weights of the objects. Human postures and their motions are automatically recomputed by the simulation software to reflect the updated scene. Following the successful execution of the ergonomic analysis of the given task, an ergonomic report and time estimates are generated by the system.

In this study, two digital human models with equal physical features, a shelf and a matrix of bottles were modeled in the simulation environment. To simulate the conditions of a supermarket, the human models were programmed with the task of stacking bottles in the shelves – one was assigned with the role of a giver and the other as a receiver (Fig. 1a). Only one bottle was handed over at a time (Fig. 1b). The shelves in the simulation environment had the dimensions, 145 cm – upper shelf height, 95 cm – medium shelf height and 43 cm – lower shelf height. The weight of the bottles, the height of the shelves (higher/medium/lower) and the physical attributes of the human model (body mass index, BMI) were varied and its influence on the fatigue measures was evaluated.

| Max pressure on the Lower Back in Newton (GIVER) | | | | Max pressure on the Lower Back in Newton (RECEIVER) | | | |
|---|---|---|---|---|---|---|---|
| High | Mid | Low | Shelf Height/%tile of Population | High | Mid | Low | Shelf Height/%tile of Population |
| 1855.21 | 1896.63 | 2011.15 | 5 | 1194.28 | 1300.51 | 1883.62 | 5 |
| 2602.33 | 2610.38 | 2602.44 | 50 | 1598.45 | 2484.55 | 2399.33 | 50 |
| 3220.87 | 3309.40 | **3411.50** | 95 | 2070.48 | 2186.89 | 3149.42 | 95 |

| Average pressure on the Lower Back in Newton (GIVER) | | | | Average pressure on the Lower Back in Newton (RECEIVER) | | | |
|---|---|---|---|---|---|---|---|
| High | Mid | Low | Shelf Height/%tile of Population | High | Mid | Low | Shelf Height/%tile of Population |
| 1276.68 | 1313.21 | 1318.58 | 5 | 453.47 | 471.55 | 788.07 | 5 |
| 1667.48 | 1664.92 | 1692.25 | 50 | 634.80 | 794.12 | 999.25 | 50 |
| 2203.01 | 2256.09 | 2245.90 | 95 | 744.52 | 725.14 | 1302.57 | 95 |

**TABLE 2** Results of the software simulation in Jack 8.0.1. Output of the Lower Back Analysis (LBA) with the average and maximum pressure sustained on the lower back for the Giver [Left] and Receiver [Right] for different shelf height (Low/Mid/High) and for different population group (5/50/95)

The analysis was done for three groups representing 5, 50 and 95-percentile of the population with BMI equal to 23.5, 25.2 and 27.7 respectively [41], [42]. This population distribution and its physical characteristics is an in-built feature of the simulation software and is based on the 1988 US Army Anthropometric survey (ANSUR) [42].

The range of stress and fatigue for each of the individual roles was derived from the simulation. The fatigue measures included Lower Back Analysis (LBA), Estimated Recovery Time Needed, and Muscle Strain Time History. The LBA metric uses a complex biomechanical low back model to evaluate the spinal forces that act on the lower back for any posture and loading conditions [40].

### 4.1 Software Simulation Results

Simulation results are presented in detail in Table 2. It shows the output of the Lower Back Analysis (LBA) indicating the average and maximum pressure (in Newton) sustained on the lower back for the giver and receiver for different shelf height (Low/Mid/High) and for different population group (5/50/95 percentile).

Fig. 2 shows the Lower Back Analysis (LBA) of the giver (in blue-dashed) and the receiver (in red-bold) belonging to the 95-percentile population group during the course of the task of shelving bottles in the higher shelf.

Figure 2 indicates that the peak to peak difference, $P_{Peak-Diff}$, in the sustained pressure (which is the difference between the variable's extreme values) between a giver and a receiver is 36%. The average pressure sustained by the giver is 2203 Newton *(SD=825)* and by the receiver is 774 Newton *(SD=210)*, resulting an avg. difference $P_{Avg-Diff}$ of 65%.

## 5. IN-HOUSE LAB EXPERIMENTS

### 5.1 Experimental Design

*(1) The Scenario:* Figure 3a illustrates the real-life scenario of a supermarket that was re-created inside the IMT Robotics Lab of Ben-Gurion University of the Negev (BGU). The experimental area, as shown in Fig. 3b, consisted of an empty shelf and a set of 120 soft-drink bottles (filled with water) of 1.5 liters each weighing approximately 1.5 kg. The bottles were arranged in the form of a matrix of dimension 5x24 on the floor, in front of the shelf (45 cm away). The shelves used in the experiment were approximately of same dimensions as found in supermarkets (165 cm – upper shelf height, 124 cm –

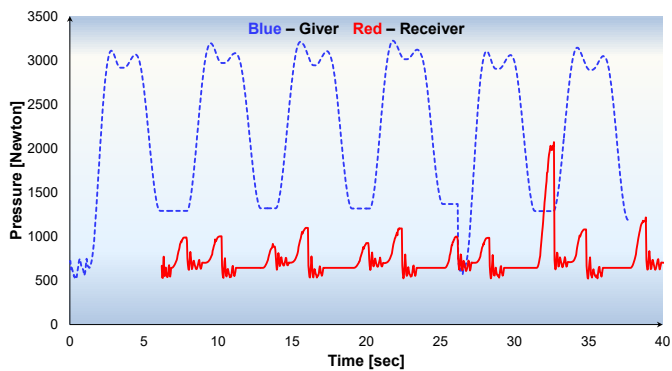

**Fig. 2** The Lower Back Analysis (LBA) for a handover task (higher shelf & 95-percentile population) [Giver – Blue dashed & Receiver – Red bold line]

medium shelf height and 10 cm – lower shelf height). The given task was to fill the empty shelf with these bottles.

*(2) Conditions:* The experiment included two variables – shelf height (Higher Shelf/Lower Shelf) and frequency of handover (Normal Mode/Competitive Mode). In the normal mode, the teams were expected to work at normal pace without any time pressure or productivity target. In the competitive mode, teams were noted to work faster than the normal mode. The motivation behind the competitive mode was to re-create the peak hours/days of a supermarket prior to weekends, holiday seasons when the work pressure mounts up and expected output increases considerably.

Subjects were informed that the team with the highest throughput in competitive mode will receive a prize. However, in either of the modes, no productivity target or time pressure was given. Furthermore, in both the modes subjects were unaware of the total time for which they are supposed to carry out the given task in each phase, the total number of bottles to be placed in the shelf and the total number of bottles in the inventory. They were informed that they need to continue until they were asked to stop. The motivation behind this was to ensure they invest their energy smartly so that they are able to work for longer stretches of time. The goal was not to empty the inventory which is considered to be countless, but rather to ensure a natural speed of work to resemble the speed of supermarket workers who do this job for a span of 8 hours.

The experiment had four phases and each of the pairs went through all four conditions: (i) Normal Mode – Higher Shelf (ii) Normal Mode – Lower Shelf (iii) Competitive Mode – Lower Shelf (iv) Competitive Mode – Higher Shelf. The shelf height was assigned randomly; however, the competitive mode followed only after the Normal mode was executed for both the shelves. This was done to:

*(a)* ensure that subjects indeed work at normal pace during normal mode so as to save energy for the competitive mode which would be towards the end of the experiment.

*(b)* avoid the risk of collecting erroneous data during normal mode if during the transition from competitive to normal mode, subjects continue to remain faster than their usual pace.

### 5.2 Experimental Procedure

The participants received experimental instructions in written form and then entered the experimental arena, where the experimenter orally provided the task instructions in further detail. Subjects were asked to visualize the arena as a supermarket and themselves as its full-time workers. They were aware that the experiment is video recorded.

Subjects were then given the opportunity to ask any clarifying questions before the experimenter left the participants to start the given task. They decided mutually among themselves on-the-spot the roles of giver and receiver and were instructed that the role they choose to opt must remain same during the entire course of the experiment. We intentionally did not pre-assign and impose upon the role because it was clear that the job of a giver was apparently more difficult than the job of a receiver. In addition, we were interested to see how the negotiation between the subjects takes place during a joint-action.

The rules of the given task were – (a) receiver is not allowed to pick the bottle directly from the floor. He/she can collect the bottle only from the hands of a giver (b) The giver cannot place the bottle directly on the shelf; (s)he can only deliver the bottle to the receiver (c) A giver can pick any bottle from the given stack of 120 bottles. Subjects were however, not aware of this exact number 120 - the total number of bottles that is in the inventory.

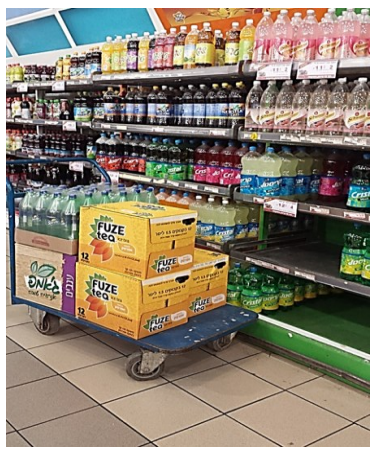 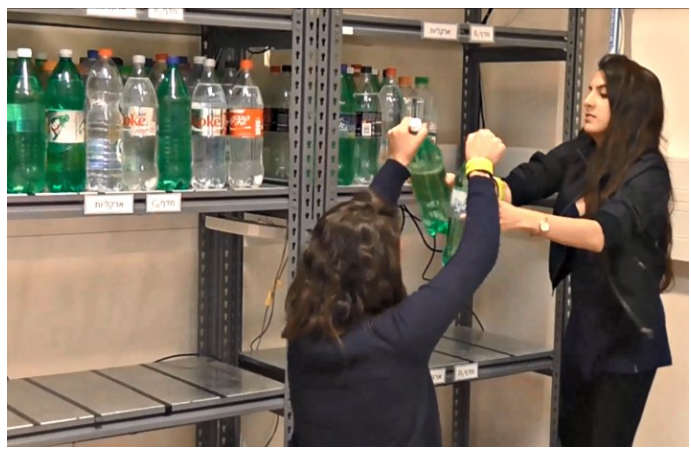

**Fig. 3a** Field-studies were done in this specific area of the supermarket; Bottles are in the cartons stacked at two levels (Upper Carton / Lower Carton)
**Fig. 3b** The Experimental Arena

(d) A giver can pick up the bottle in any physical posture he feels comfortable (e) The receiver is expected to align the bottles in the shelf elegantly as seen in a supermarket. If the shelf is visualized as a matrix, then each column must have at least 4 bottles (preferably 5) and each row with at least 9 bottles (preferably 10).

No other rules were defined with the aim to enable the subjects to behave and evolve naturally during the course of the task. Intentionally, the number of bottles the giver can pick up at a time during each handover cycle was not defined to observe how it evolves naturally.

The subjects worked for 2 minutes in each of the four phases and were allotted a break of 4 minutes after each phase. Break sessions were kept to reduce the impact of fatigue following each phases as much as possible. Each of the phases concluded in the form of external interruption ("Time Up") by the experimenter since the time duration of each phase was unknown to the subjects. They were free to talk and interact among themselves during the course of the task; however, during the break they were asked to sit at two different locations away from each other to regulate any possible communication about their experiences during break session.

### 5.3 Dependent Measures

*Task-related performance* of the participants was measured by objective analysis of the number of bottles shelved every 10 sec during each phase; and, through off-line video data analysis of the joint-task, focusing on the level of coordination in the team during the task. The level of coordination was assessed by measuring the waiting time of the partner in every handover.

The number of bottles shelved every 10 sec measures average throughput of the given team. For each phase of the experiment, 12 measurements were taken (12x10=120sec; resulting in 2 minutes of each phase) which were analyzed to examine the average productivity. This process follows the central limit theorem in which the 12 random variables (each obeying Poisson distribution) are distributed normally, regardless of the underlying distribution.

Subjective analysis of the experiment participants was achieved through two questionnaires given during and after the experiment. During the intervals, subjects were asked to report their experiences by ranking relatively the current phase with the previous ones in terms of (i) comfort and, (ii) coordination / synchronization. While an accurate grading of their experiences in terms of comfort and coordination can be done only at the end of the entire experiment, there are good chances of error in judgement because one may not distinctly remember (or confuse) their experiences of each phase in the end. To avoid any such error, we tried to capture it as fresh as possible and hence asked the subjects to relatively rank the current phase with the previous ones during every interval.

Subjects were reminded that they are free to rearrange the ranking order as and when they feel the need.

At the end of the experiment, participants were given a post-experimental questionnaire to fill-in. The experimenter explicitly asked the participants to not rush with the questionnaire and answer the questions sincerely. This was done to ensure accuracy in data collection. In the end, subjects were given a global overview and main objective of the project.

### 5.4 Participants

A total of 42 participants (18 female, 24 male) took part in the experiment. The subjects included a mix of local and international students of Ben-Gurion University of the Negev in the age group, 21 – 35 ($M= 26.07$ years, $SD =2.29$). The weight of the participating male are ($M= 74kgs$, $SD =7.37$) and female are ($M= 55kgs$, $SD =7.79$). Subjects were asked to fill-up a health declaration form to ensure they are fit to do the job and do not suffer from health problems (joint problems, etc.) which may prevent or make it difficult for them to accomplish the task. Subjects were invited for the experiment in pairs chosen randomly to ensure that the subjects do not know each other (or may know just by face) beforehand.

## 6. EXPERIMENTAL RESULTS

### 6.1 Objective Evaluation

Table 3 shows the average productivity (nr. of bottles/10 sec) and variance of the different groups in the competitive mode (first value in each cell) and normal mode (second value in each cell). The performance of all the 21 teams for the lower and higher shelf is presented in the fourth and fifth row respectively. Results show that the average productivity in any shelf during the competitive mode and normal mode is approximately 8 and 5 bottles per 10 sec respectively. The F-Test of equality of variances was done for different combinations of shelf height and speed/mode. All the left and right ended curly brackets as indicated in Table 3 indicate a significance of $p<0.0001$.

The data was further analyzed and the teams were classified into three categories – *High-, Average- and Low-Yield Teams* – based on their productivity in the competitive mode. Teams with an average productivity rate of 9 and above were categorized as high-yield (7 Teams), between 8 and 9 as average-yield (6 Teams) and finally teams with 8 or less were categorized as low-yield teams (8 Teams). The first three rows of Table 3 show the average productivity and variance of these three productivity-based categorized groups.

The F-test of equality of variance was also performed between these three groups. Results (Table 3) indicate that the variance of the high-yield teams in competitive mode is 43% higher than the low-yield teams with a significance of $p<0.015$. The high-yield teams in the competitive mode are also the high-yield teams in the normal mode; their productivity being 25% higher and variance being 96% higher than the low-yield teams in normal mode (with a significance of p<0.0001). All other values connected by curly brackets in Table 3 have a significance of p<0.0001.

| Group | Average productivity/10sec Competitive Mode/ Normal Mode | Variance Competitive Mode/ Normal Mode |
|---|---|---|
| High Yield | 9.60 / 5.89 | 2.73 / 2.57 |
| Avg. Yield | 8.62 / 4.97 | 2.89 / 1.37 |
| Low Yield | 7.23 / 4.71 | 1.90 / 1.31 |
| Entire Group – Lower Shelf | 8.47 / 5.08 | 3.48 / 2.06 |
| Entire Group – Higher Shelf | 8.37 / 5.28 | 3.45 / 1.93 |

**TABLE 3** Average Productivity rate (Nr. of Bottles/10sec) and Variance for Competitive Mode and Normal Mode. All values connected by curly brackets have a significance of p<0.0001 (others are not statistically significant)

## 6.2 Subjective Evaluation

### 6.2.1 Team communication during task

Results (Fig. 4) show that the most frequent dialogues (39%) during the task were concerning their relative positions to each other. Example of such form of interactions include, *"I stand here, and you go there"*. Other frequent exchanges (34%) were about the number of bottles to be transferred in subsequent handover. The third most frequent dialogues were concerning the 'how to arrange the bottles in the shelf'. Together, they constituted 88% of all the different communication types. Hence, team communication during the course of the experiment was primarily related to the team strategy and mutual coordination.

### 6.2.2 Perception of the relative level of difficulty of team partner's task

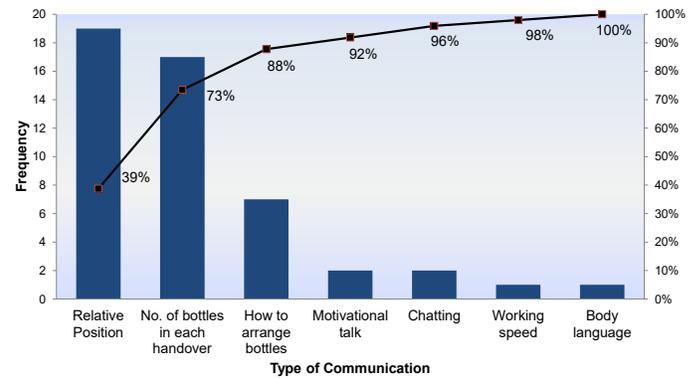

**Fig. 4** The Pareto chart of the type of communication between the team partners during the task

Results (Fig. 5a) show that 95% of the receivers rated the relative level of difficulty of the giver's task as same or as more difficult than theirs. However, only 24% of the givers say the receiver's task is "less difficult or easier" than theirs. In fact, more than 70% of the givers are calling receiver's task as same as theirs.

### 6.2.3 How the point-of-handover (p-o-h) was decided?

Subjects were asked in the questionnaire how the point-of-handover (p-o-h) was decided among them. Results (Fig. 6a) show a difference in opinion among the givers and receivers. While 48% of the givers say they decided the subsequent p-o-h by looking at the location of the team partner's hand, an equal number of receivers did not bother to give a serious thought about the p-o-h because their perception of handover is *"it happened automatically, with no thinking"*. Also, 33% of both the givers and receivers say they made the decision as per the previous handover.

Fig. 6b is a screenshot captured from the video recording which shows an instance of a giver deciding the p-o-h by staring at the location of the team partner's hand while the receiver collecting the bottle without looking towards the giver.

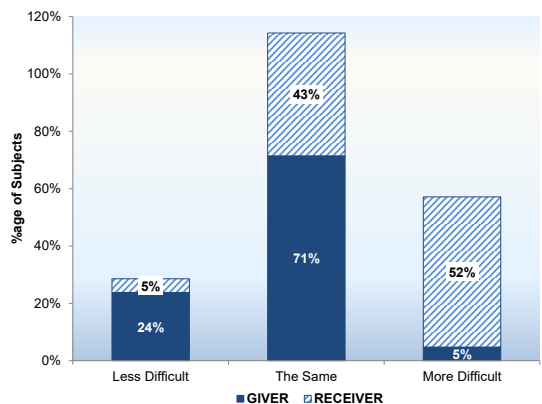
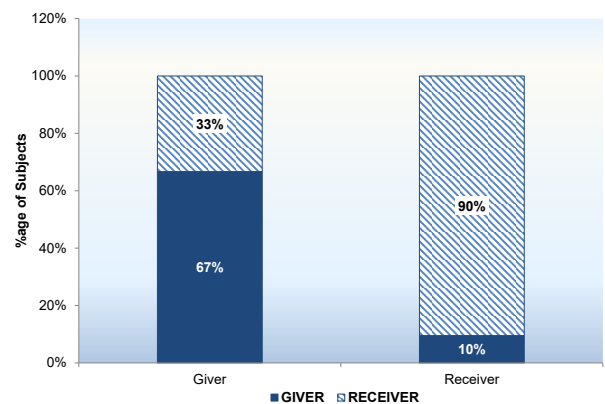

**Fig. 5a** Graph on the left showing the perception of the relative Level of Difficulty of team partner's task when compared to oneself;
**Fig. 5b** Graph on the right showing subjects' preference for future roles if given a choice

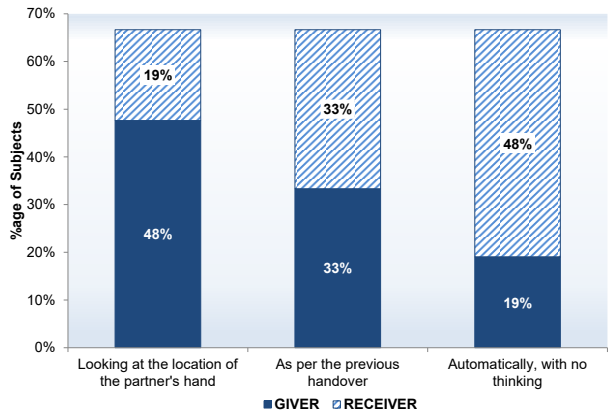 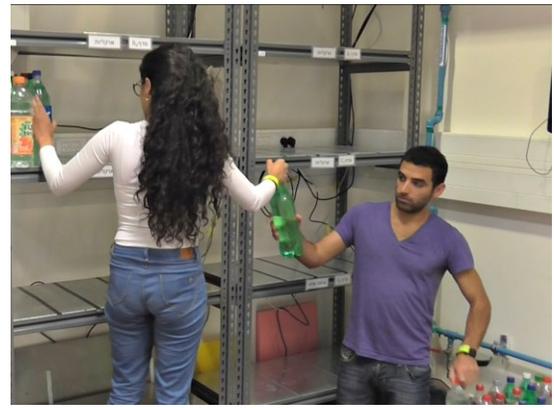

**Fig. 6a** How the decision of the point-of-handover was made during the task; **Fig. 6b** A giver deciding the p-o-h by staring at the location of the team partner's hand while the receiver collecting the bottle without looking towards the giver.

*6.2.4 Habit persistence in Joint-action?*

Results (Fig. 5b) indicate that subjects, in general, tend to stick to their current roles if given a choice to switch (their roles) in the future experiment. 66% of the givers prefer to remain as givers and 90% of the receivers prefer to remain as receivers in future roles. Given that the job of a giver is way more difficult than that of a receiver, a comparison of the relative difference in the response of the givers (66%) and receivers (90%) is not appropriate.

*6.2.5 Perception of the level of commitment of the partner*

Results (Table 4, Q5; Appendix) show that 95% of the team members rated their partner (givers rating receivers and vice-versa) as equally committed during the task.

*6.2.6 Relative ranking of the most comfortable phase and the most well-coordinated phase of the experiment*

Results (Fig. 7) show that 62% of all the subjects generally felt most comfortable working in the normal mode. Lower shelf-normal mode was chosen as the favorite by the majority (67%) of the givers (Fig. 7a) and higher shelf-normal mode was chosen by 57% of the receivers (Fig. 7b).

The relative ranking of the best level of coordination (Fig. 7), results indicate that 30% (highest among the other options) of the givers rate lower shelf-competitive mode as their favorites while 43% (highest among the other options) of the receiver rate higher shelf-competitive mode as their favorites. In general, however, givers showed more diversity in choosing their favorites for the best level of coordination as compared to receivers.

*6.2.7 Relative speed and rhythm perception of the team partner and how they adapted to each other*

Results (Table 4, Q3 & Q4; Appendix) indicate that most of the subjects (95%) noted they developed a rhythm and that they adapted themselves to match the speed of their partner. When asked about the perception of the relative speed of the partner compared to themselves, Fig. 8 shows that 64% of all the subjects (givers & receivers together) considered their partners' speed as inconsistent (i.e., sometime fast/sometimes slow).

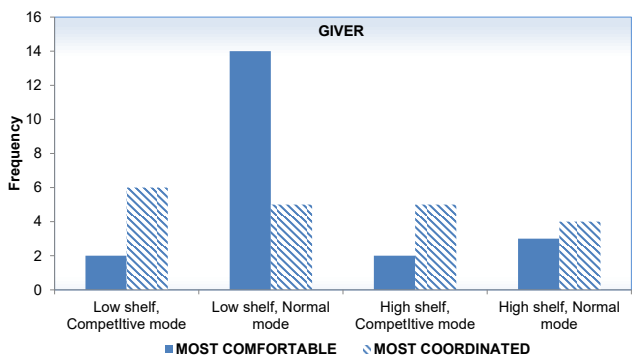 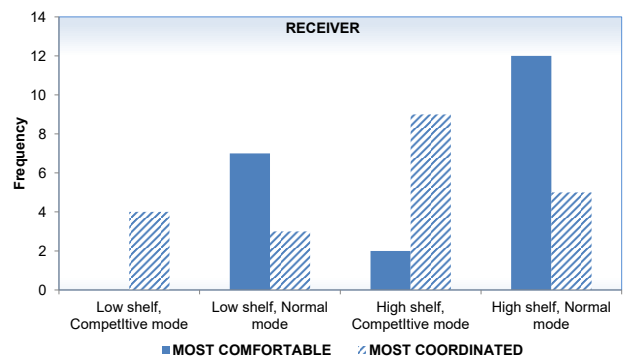

**Fig. 7a & 7b** Relative ranking of the most comfortable phase and the most well-coordinated phase of the experiment for Giver (left) and Receiver (right)

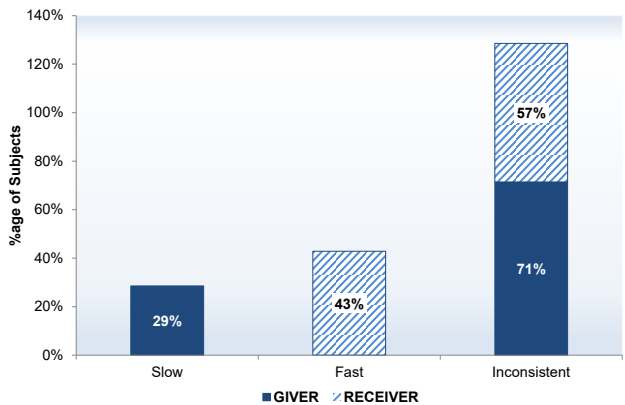

**Fig. 8** Perception of the speed of the team partner

In addition, the proportion of givers perceiving the receiver as 'slower' is significantly higher ($p<0.005$) than the proportion of receivers perceiving the giver as 'slower'. Similarly, the proportion of receivers perceiving the giver as 'faster' is significantly higher ($p<0.0001$) than the proportion of givers perceiving the receiver as 'faster'.

### 6.2.8 Preference for using two hands (2 bottles at a time)

Results (Table 4, Q2; Appendix) show that, if given a choice, 76% of the giver and receiver prefer to transfer 2 bottles at a time as compared to one.

### 6.2.9 Subjects' preferences towards working together

Results (Table 4, Q6; Appendix) show that 90% of the subjects prefer to work in teams even if that means they need to do double the work as a whole (100 bottles together) as compared to the option of shelving 50 bottles alone.

### 6.2.10 Tendency when team coordination was perceived as perfect

Results (Table 4, Q7; Appendix) show that no subjects (0%) felt the urge to slow down when the subjects perceived that the coordination between them and their partner was going perfect. They either felt the tendency to speed-up (60% of the subjects) or maintain that speed (40% of the subjects).

### 6.2.11 Fatigue

Results (Table 4, Q8; Appendix) show that almost 85% of the subjects felt a level of fatigue that is equivalent to 'not at all' or 'a bit tiring'. Also, 95% of the subjects mentioned that the allotted break time between each mode was enough to recuperate from the tiredness.

### 6.3 Off-line Video Analysis

A clip of the video recording of the in-house lab experiment study is presented in Online Resource 2. The clip gives a better understanding of the experiment design and the impromptu handover strategies used by the subjects.

### 6.3.1 Point-of-Handover (p-o-h)

Video analysis shows that there is not always an eye-gaze between the subjects or at the p-o-h in every handover cycle (see Online Resource 2). In these cases, the giver reaches up to the receiver's hand while the receiver extending his/her hand unconsciously to the previous p-o-h and keeping his/her gaze fixed towards the shelf.

### 6.3.2 Location of the grasps on the bottle

Video analyses indicated that when the subjects were delivering two bottles at a time, they tend to grasp the neck of the bottles. The tendency increases further when the bottle is transferred to the lower-shelf two at a time. Whereas, when the handover was one at a time, subjects tend to grasp the body of the bottle. More so, when the bottle is transferred to the higher-shelf and one at a time, it is observed that subjects tend to grasp the body of the bottle.

### 6.3.3 Negotiation in choosing roles of the joint-task

Video analysis of the experiment shows that subjects mutually decided among themselves the role they want to choose and never had any disagreement during the negotiation; the negotiation was approximated to be achieved in less than 20 sec. The negotiation took place mostly non-verbally.

### 6.3.4 Task expertise

Video analysis indicated that the subjects slowly developed some sort of understanding in terms of team strategy and mutual speed of working which resulted in better action coordination over time. This however did not remain constant and kept evolving depending upon the given experimental conditions (shelf height and mode).

## 7. DISCUSSION

The contribution of this article is a systematic approach to describe human-human joint-action in short-cycle repetitive tasks in a quantitative and qualitative way.

### 7.1 Objective Analysis

Analyzing the values in the horizontal rows from left to right (Table 2) indicate that the pressure exerted on the lower back is always higher for a lower shelf as compared to a higher shelf. The results in Table 2 also indicate that for all the combinations of shelf height and population distribution simulated in the study, the pressure exerted on the lower back of a giver is always higher than that of a receiver irrespective of the shelf height. The maximum pressure on the lower back of a giver belonging to the 95-percentile population group may go up to 3411.50 Newton (Table 2). According to [43], this pressure crosses the threshold safety-limit for lower back. Overall, it clearly indicates that **the task of the giver is physically more strenuous than that of the receiver.** Surveys on the supermarket workers during the field studies also indicate the same. Statistical analysis of the work-methods field studies data (Table 1) indicate that there is an influence ($p<0.1$) of the difference between 'bottle carton height' and 'shelf-height' on avg. cycle times. The lowest difference accounted for faster cycles times and highest team performance. This observation could be explained using simulation study which shows that the posture, in this case, requires the least amount of relative effort and offer better ergonomics to the giver and the receiver.

The average productivity rate in the experiment was found similar to what was measured in supermarkets confirming that the experimental data was collected in close to real conditions.

The objective analysis of the experimental data (Table 3) indicate that the high-yield teams have the highest variance in productivity every 10 sec (43% higher in competitive mode, $p<0.015$ and 96% higher in normal mode, $p<0.0001$) as compared to the low-yield teams. So, the teams achieved higher productivity at the cost of higher variability in the handover frequency. Lower value of variance among the low-yield teams indicate that the variability in handover frequency was lower which, in turn, means the handover cycles were mostly stable indicating good coordination. *This implies the "low-yield teams" in terms of productivity are, in fact, the most well-coordinated teams.*

When the productivity of these three performance-based groups was checked for normal mode, it turns out that the high-yield and low-yield teams in competitive mode continue to remain high-yield and low-yield teams in normal mode respectively. Variance of the high-yield teams is also significantly higher (96%, $p<0.0001$) than the other. **This implies that the art of well-coordination among the team partners is probably not influenced by the increasing or decreasing frequency of handover.** The well-coordinated teams continue to remain well-coordinated in either of the modes and vice-versa.

The existence of higher variability in high-yield teams can be explained based on the speed-accuracy trade-off of motor control in prehensile movements which follows Fitts' Law [38]. As the speed of the team increased, the point of handover kept moving (from L to R or from R to L along the length of the shelf) at a faster rate which might influence the variability following Fitts' Law. Since, increased variability resulted in poor-coordination, 'accuracy' is in a sense compromised here.

The variability in handover frequency may also be attributed to the evolving team dynamics and bottle placement strategy. In other words, the differences in the way the bottles are placed in the shelf and the number of bottles delivered in each handover may have an effect on the variability. Bottle placement strategy was not used as a metric to measure variability because different teams came up with different strategies (each individual team sometimes even had different strategies for upper / lower shelf). As a result, it could be hard to classify this metric into a small *'n'* number of categories to understand its role on productivity.

In addition, as observed during offline video analysis, when the expertise in handover grows over time during the course of the experiment and participants are starting to develop some sort of coordination between them, variability tends to decrease. However, based on section 6.2.11, it can be said that there was no significant impact on the variability due to fatigue.

### 7.2 Subjective Analysis

#### 7.2.1 Team Communication

The communication among the team partners mostly consisted of interactions related to team strategy and their relative position to each other. Researches in sports psychology show that when team members are in situation where verbal communication is feasible (in terms of physical distance, time taken to communicate) then, inexperienced teams prefer to communicate task–specific knowledge through intentional verbal communication to ensure high-level of accuracy of the message transferred [24], [25], [44], [45]. Considering that the subjects were doing the given job for the first time, the observation can be related to the above explanation.

#### 7.2.2 Modesty and 'We-Mode' influences Team dynamics

**Almost every subject considered their team partner as equally committed towards the task.** This probably influenced the way subjects rated the relative difficulty level of the task of the other (Fig. 5a). This observation can be concluded, in other words, as both partners in more than 90% of the teams show a certain level of modesty when rating their partner's performance. Support for this result can be found in Applied Psychology research where it has been shown that at an individual level of analysis, personality traits like agreeableness have a major influence on *peer-ratings* of team member performances [26] irrespective of job-specific skills and general cognitive ability [27]. Probably, there existed a good level of agreeableness (the tendency to be good-natured, cooperative, and trusting) between the partners of our experiment. This can be checked from the answer to the question where subjects were asked to rate the level of commitment of their team partner. Results show that 95% of the team members rated their partner as equally committed during the task.

Another way to explain this observation is based on the judgement of control over joint-action. For the success of the given collaborative task, optimal team-performance was the key. Since there was a common goal for both the partners, in effect, there was a congruency between each partner's intentions and actual action which influenced positively in their judgement of control. This probably developed a spirit of "we-mode" [46] among the team partners where the common goal becomes more important than individual needs and preferences. So, we may say modesty and 'we-mode' influences team dynamics in joint-action.

*7.2.3 Habit persistence in Joint-action*
From 6.2.4, we see that **even though the job of a giver is apparently and ergonomically more difficult than that of a receiver, majority (two-thirds) of the giver opted to stick to their current role** (Fig. 5b). This observation is generally explained by psychologists using two types of theories.

One is the theory of habit persistence in decision-making [28], [47], [48]. According to this theory, habit plays a certain amount of role in decision-making. In our experiment, subjects spent only 20 minutes on the task as a whole (including break sessions) and even then, there exists clear signs of habit persistence.

The other way to explain this observation is based on the theory of motor preservation. It says that human tend to continue to make the same movements by recalling, evaluating and re-generating stored postures instead of re-planning during movement planning (especially in reaching tasks) [49], [50].

*7.2.4 Emergent Coordination*
**Subjects on one hand were inconsistent (Fig. 8) and on the other were rhythmic**. This observation of adapting themselves to form a mutual rhythm is termed as emergent coordination [51] by the psychologists where the partners sometimes speed-up or slows down depending upon the context to match their partners speed giving rise to a rhythm between the partners.

*7.2.5 Leading and Lagging*
Analyzing the results of the subjects' perception of the speed of the other (Fig. 8), it can be said that givers were generally perceived as faster than receivers and vice-versa. Considering, movement synchronization is a guiding dynamical process which leads to stable coordination patterns in natural human-human joint-action, it can be concluded that **givers led and set the pace of coordination.**

*7.2.6 Does most comfortable/ergonomic work method when speeded up apparently gives a perception of most well-coordinated joint-action?*
Subjects, in general, felt most comfortable working in the normal mode. Such preferences can be easily explained on the basis of minimum bio-mechanical efforts and strain that one has to put-in for their chosen favorites. The simulation study also shows that these modes offer least fatigue and better ergonomics for their respective roles.

The relative ranking of best-coordination, however, show that there exists a possible trend. **Subjects perceive the competitive mode of their most comfortable working mode respectively, as the most well-coordinated phase of the experiment**.

The objective analysis (section 7.1) indicated that the well-coordinated teams have the least variability in handover cycles in both competitive and normal mode. However, based on the above-mentioned trend, it can be concluded that even in a well–coordinated team, each of the subjects may perceive the period of most well-coordinated joint-action at different times. This is probably because **the act of perceiving the level of coordination among team partners may not be the same for the same joint-action** as it does not depend solely on the variability in handover cycles but also upon their perceived effort which varies depending on the subject's role (giver or receiver).

*7.2.7 Preference towards the use of two hands*
From 6.2.8, it can be concluded that **subjects, in general, have a preference towards using two hands together for the given task**. This observation is generally explained by theories in motor control and bimanual coordination [52]–[55].

*7.2.8 Negotiation in Decision Making*
A high level of agreeableness between the partners is probably the reason for such frictionless negotiation among team partners for choosing their individual roles (from 6.3.3). The agreeableness factor has its roots in applied psychology [27] and has been explained above in section 7.2.2.

*7.2.9 Preference towards working in Teams*
Subjects have shown clear **preference for working in teams even if that requires them to work more** as compared to working alone (from 6.2.9).

*7.2.10 Rhythms that speed us up*
Support for the behavioral tendency of the subject when the team coordination was perceived as perfect (section 6.2.10) can be found in the researches in Musicology and Psychology where it has been shown that **humans feel the urge to speed-up under certain rhythms** [56]. The current observation goes in line with these findings and has implication for adaptive control system design which is discussed in the next section.

*7.2.11 Point-of-Handover*
The current research looks into the anticipation of the p-o-h of the subsequent handover in short-cycle repetitive task based on the experience of previous handovers. Results (section 6.2.3) show that the decision on handover point is taken sub-consciously or automatically by the giver/receiver in many of the cases. In other cases, both receiver and giver expect the handover to take place around the same location as the previous handover.

It is to be noted that the handover in our experiment is facilitated *without* necessarily having an eye-contact between the giver and receiver (from section 6.3.1). So the p-o-h is not necessarily being decided by the giver as in [19] but it varies in this experiment. For example, when the giver and receiver do not have an eye-contact, then the p-o-h is unconsciously decided by the **receiver** as the giver reaches up to the receiver's hands (while the receiver extending his/her hand unconsciously to the previous p-o-h and keeping his/her gaze fixed towards the shelf). This observation supports previous findings that in repetitive handovers, **a-priori expectation of the receiver about the probable p-o-h plays an important role in the success of the handover** [19].

*7.2.12 Location of the grasps on the bottles*
Results of section 6.3.2 can be explained based on the end-state comfort effect governing motor control which predicts that "people will grasp an object for transport in a way that allows joints to be in mid-range at the end of the transport" [57], [58]. It also supports the findings of [59], [60] that it is probably a distinct effect of *recall and generation* on movement planning and that "end-state comfort effect facilitates joint-action" [61].

## 8. CONCLUSIONS

Overall, this work gives us a better understanding of the role of a giver and a receiver in a short-cycle repetitive handover task. The subjective analysis showcases *individual differences in the way a common joint-action is being perceived by team partners depending upon their role.* The conclusions presented below are in the form of recommendations for designing human-robot collaborative systems for short-cycle repetitive tasks. They are derived from the implications of the discussion on the analyses of human-human joint-actions. Each of the recommendations is related to a specific derived source in section 7 and indicated if it is a direct or indirect implication.

1. The job of a giver is physically more strenuous than that of a receiver. Thus, **the job of a giver should ideally be replaced by a robot** if an H-R system is commissioned in a supermarket [based on section 7.1].

2. Humans, in general, prefer to stick to their roles and habits [directly implied from 7.2.3]. This probably means, they would also prefer to stick to their own convenient pace of working. So, in general, **the collaborative robot should be able to learn the preferences of the user to offer more personalized service in subsequent collaborative task**. An option to save the preferences and profile of each user in robot's database is thus recommended. [62] shows how a robot-human handover is improved based on human preference feedback.

3. Any H-R system commissioned for such tasks should have the possibility to play both the role of a giver and a receiver to be able to work in any given role depending upon the choice of the user [directly implied from 7.2.3]. While, a single arm robot with a mechanical double bottle holder may be able to accomplish the job of a giver, it will not be able to play the role of a dual-arm receiver which also requires aligning the bottles in the shelf. Hence, **a dual arm collaborative robot** (like ABB Yumi and Baxter) **is better suited** to perform both of these roles in equal capacity when collaborating with a human [directly implied from 7.2.7].

4. Humans prefer to work in normal mode in comparison to competitive mode [directly implied from 7.2.6]. So, **the default speed settings of the collaborative robot should be the average working speed of human**. This will also reduce fatigue and stress.

5. Humans tend to be well-coordinated when they are in the competitive mode of the most comfortable posture/work method which, in turn, is dependent upon the chosen role and shelf-height. The vice-versa is also true [directly implied from 7.2.6]. Therefore, the minimum and maximum speed of each handover cycle of a collaborative robot for different modes, roles and shelf heights can be defined and calibrated using this principle.

6. **A high degree of agreeableness between the team partners of a joint-task is very important** to fruitful and mutually satisfying work experience [directly implied from 7.2.2]. Hence, collaborative robots must be social (e.g., Baxter).

7. A-priori expectation of the receiver about the probable p-o-h plays an important role in the success of the handover in short-cycle repetitive task [directly implied from 7.2.11]. Well-coordinated teams continue to remain well-coordinated in all the tested conditions and hence, they have the least variability in their handover frequency [directly implied from 7.1].

Now, in the case of short-cycle repetitive tasks in supermarkets where duration of each handover cycle could be as low as 2-3 seconds, generating 100% accurate adaptive motion and determining the exact p-o-h for each handover cycle through action coordination based on Keller's framework [7] [29], of adaptation, attention and anticipation will generate non-rhythmic motions (due to processing times involved in delivering high accuracy) resulting in a stop-and-go motion with no fluency in joint-action. This may potentially have a high cost on team coordination and productivity.

Based on the findings of sec. 7.1 and 7.2.11, we argue that if team productivity is deemed critical for a human-robot system executing a short-cycle repetitive task, **a robot with a fixed periodic motion and a fixed p-o-h preset by the respective user is probably better suited than highly accurate systems with non-rhythmic or reactive motions.** This is because, a robot working with a fixed rhythm is more well-coordinated and predictable than any other system. So, mutual coordination can be achieved easily through human adaptation because humans are considered experts in working jointly in rhythmic activities. A recent study has also shown that human adaptation in a human-robot system can significantly improve team collaboration [63].

The robot should however be equipped with advanced sensors to be able to track the human partner as a whole for valid safety reasons. Also, the robot must be able to understand that the job is over / at pause and should not blindly continue the fixed rhythmic motion for indefinite time.

This type of fixed rhythmic robot motion is similar to the pro-active behavior as demonstrated in a human-robot repetitive handover experiment reported in [17]. Results as reported in [17] show that the pro-active method provided the greatest levels of team performance but offered the poorest user-experience compared to the reactive and adaptive methods. The reactive motion offered the best user-experience but at the cost of the lowest level of team performance while the adaptive motion offered a balance of the two requirements.

In a fixed rhythmic H-R interaction, the user should ideally be given the opportunity to pre-define the robot's periodic motion and the p-o-h during learning by demonstration phase to ensure that the user is very much in control of the desired speed and p-o-h. This could offer, somewhat, a better user-experience and may offset the poor user-experience involved in such pro-active behavior. It can be said based on a recent study [64] which shows that the possibility to customize the interaction with a robot as per one's individual preferences creates a sense of "self-agency" in humans which have a strong positive influence on user-experience.

## 9. LIMITATIONS OF THE APPROACH AND SCOPE FOR FUTURE WORK

The current work focuses on the subjective experience and psychological perspective of givers and receivers in short-cycle repetitive handover tasks. The objective measures presented here are limited in nature because others are related to the bio-mechanics, work-methods and ergonomic aspects of the given task which is not the focus of this paper. As a result, the objective and the subjective measures could not be contrasted for bias or consistency.

Having said that, the work-methods field studies has been instrumental in designing the set-up of the lab experiments and most importantly, towards defining the experiment variables and conditions. Similarly, the simulation study gave insights on the ergonomic aspects of the role of givers and receivers which also influenced the design of the lab experiment (for example, it helped us decide that the alignment of giver (G), receiver (R) and the matrix of bottles (B) should be R-B-G and not R-G-B. That is, when the bottles are kept in between the giver and receiver, it is comparatively less tiring for the giver.

The simulation software used in this work does not offer the flexibility of modeling the actual movements recorded during the lab experiment. Using other software that offers such possibility may help to characterize the variability of energy according to different techniques/experience/body shapes etc.

Future work should include a more detailed investigation of the actual movements from video analysis or by motion tracking with passive markers or wearable sensors which, can then be contrasted with the subjective measures of this experiment to get a broader understanding of human-human joint-action in the areas investigated in this research.

The findings of the current study should be implemented and validated in real human-robot systems executing a short-cycle repetitive task in physical space to be able to develop comprehensive recommendations and design guidelines for human-robot collaborative systems.

## 10. ACKNOWLEDGEMENTS

This manuscript is a prior version of the article accepted in the International Journal of Social Robotics. The supplementary online resources mentioned in the paper can be found in Springer website. The final publication is available at Springer via DOI http://dx.doi.org/10.1007/s12369-017-0424-9.

This research was supported by the EU funded Initial Training Network (ITN) in the Marie Skłodowska-Curie Actions (MSCA) People Programme (FP7): INTRO (INTeractive RObotics research network), grant agreement number: 238486 research project and partially supported by the Helmsley Charitable Trust through the Agricultural, Biological and Cognitive Robotics Initiative, the Marcus Endowment Fund and the Rabbi W. Gunther Plaut Chair in Manufacturing Engineering, all at Ben-Gurion University of the Negev (BGU).

The authors acknowledge their thanks to the contributions of Netta Ben Zeev, Michael Kozak, Polina Kurtser from BGU and especially to Prof. Guy Madison from Umea University, Sweden who introduced the authors to the research in rhythmic joint-action in human which led to this study. Sincere thanks also go to the anonymous reviewers who have given immensely valuable feedback.

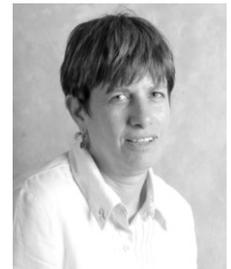

**Yael Edan** is a Professor in the Department of Industrial Engineering and Management. She holds a BSc in Computer Engineering and MSc in Agricultural Engineering, both from the Technion-Israel Institute of Technology, and a PhD in Engineering from Purdue University. Her research area is robotic and sensor performance analysis, systems engineering of robotic systems; sensor fusion, multi-robot and telerobotics control methodologies, and human-robot collaboration methods with major contributions in intelligent automation systems in agriculture.

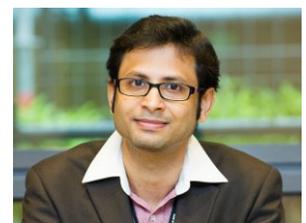

**Roy Someshwar** is currently a Qualitest Search Specialist at Google Inc. He was a E.U. Marie Curie Fellow in the Intelligent Robotics Lab of Prof. Yael Edan and Prof. Joachim Meyer (now at Tel-Aviv University) at the Ben-Gurion University of the Negev, as part of an EU-FP7-ITN project INTRO (Interactive Robotics). His doctoral dissertation is on developing robots as teammates and revolves around human-robot collaboration, joint-action and temporal coordination in human-robot teams.

He has been a visiting guest lecturer on Human-Robot Interaction (HRI) at Hochschule Karlsruhe, Germany and has worked as a Junior Scientist in the R&D of Treelogic S.L, a Spanish ICT and robotics company on the development of a commercial mobile robotic platform for HRI.

## 12. APPENDIX

| Question | Subjects' Responses | | | |
|---|---|---|---|---|
| | Giver | Receiver | | |
| Q1. Which role (Giver or Receiver) would you like to play if you have had to redo the given task? | Giver - 14<br>Receiver - 7 | Giver - 2<br>Receiver - 19 | | |
| Q2. Given an option, would you prefer shelving (A) 1 bottle at a time or (B) 2 bottles at a time? | A - 5<br>B - 16 | A - 5<br>B - 16 | | |
| | Yes | No | | |
| Q3. Do you think you and your partner developed a rhythm among yourselves slowly over the course of the cycle? | 39 | 3 | | |
| Q4. Did your partner also adapt himself to match your speed of working? | 37 | 5 | | |
| Q5. Do you think your partner was equally committed during the work? | 40 | 2 | | |
| | A | B | | |
| Q6. Given an option, which one would you choose – (A) Shelving 100 Bottles working together OR (B) shelving 50 bottles alone? | 38 | 4 | | |
| | A | B | C | |
| Q7. When the coordination between you and your partner was going perfect, did you feel the urge/natural tendency to (A) Maintain that speed (B) To Speed Up (C) To Slow Down | 16 | 25 | 0 | |
| | # 1 person did not answer this question | | | |
| | A | B | C | D |
| Q8. How would you rate your tiredness after this job? (A) Not at all (B) Bit tiring (C) quite tiring (D) very tiring | 6 | 30 | 4 | 2 |

**TABLE 4** Some of the questions with subjects' responses from the Questionnaire used in the experiment